\documentstyle[12pt]{article}

\font\tenrm=cmr10
\font\tenit=cmti10
\font\elevenbf=cmbx10 scaled\magstep 1
\font\elevenrm=cmr10 scaled\magstep 1
\font\elevenit=cmti10 scaled\magstep 1

\font\ninerm=cmr9

\newcommand{\be}{\begin{equation}}
\newcommand{\ee}{\end{equation}}
\newcommand{\bd}{\begin{displaymath}}
\newcommand{\ed}{\end{displaymath}}

\renewenvironment{thebibliography}[1]
{\elevenrm
\begin{list}{\arabic{enumi}.}
{\usecounter{enumi} \setlength{\parsep}{0pt}
\setlength{\itemsep}{3pt} \settowidth{\labelwidth}{#1.}
\sloppy
}}{\end{list}}

\newcounter{currentnumber}
\def\thecurrentnumber{\arabic{currentnumber}}
\def\eqn#1{{\rm (\thecurrentnumber.#1)}}

\def\pl#1#2{#1\kern-0.#2em\hbox{\kern-2.77pt\lower4.7pt
\hbox{$\mathchar"012C$}}\hbox to 0.#2em{}}
\def\pla{\pl{a}{01}}

\begin{document}

{\tt
``Advanced Electromagnetism: Foundations,
Theory and Applications'', eds.~T.~Barrett
and D.~Grimes, World Sci.\ Publ.\ Co, Singapore 1995,
pp.~496-505.}

\begin{center}
\vglue 0.6cm
{%{\tenbf WORLD SCIENTIFIC PUBLISHING COMPANY \\}
{\elevenbf  \vglue 10pt
NON-ABELIAN STOKES THEOREM
\\}
\vglue 1cm
{\tenrm BOGUS\L AW BRODA\footnote{
\ninerm\baselineskip=11pt 
e-mail: {\tt bobroda@mvii.uni.lodz.pl}
 }
\\}
\baselineskip 13pt
{\tenit Department of Theoretical Physics, University of
\L\'od\'z\\} 
\baselineskip 12pt
{\tenit Pomorska 149/153, PL--90-236 \L\'od\'z, Poland\\}}
\vglue 0.8cm
%{\tenrm ABSTRACT}
\end{center}
%\vglue 0.3cm
%{\rightskip=3pc
%\leftskip=3pc
%\tenrm\baselineskip=12pt
%\noindent

%}

\vglue 0.6cm
{\elevenbf\noindent 1. Introduction}
\setcounter{currentnumber}{1}
\vglue 0.4cm

\baselineskip=14pt
\elevenrm

The (standard) Stokes theorem is one of central points of
(multivariable) analysis on manifolds (see [1] for an
excellent introduction). Low-dimensional versions of this
theorem, known as the (proper) Stokes theorem, in
dimensions 1--2, and the {\elevenit Gauss} theorem, in
dimensions 2--3, respectively, are well-known and very
useful, e.g.\ in classical electrodynamics.  In fact, it is
difficult to imagine lectures on classical electrodynamics
without heavy use of the Stokes theorem.  The standard
Stokes theorem is also being called the {\elevenit Abelian}
Stokes theorem, as it applies to ordinary (i.e. Abelian)
differential forms. Classical electrodynamics is an Abelian
gauge theory (gauge fields are Abelian forms), therefore
its integral formulas are governed by the Abelian Stokes
theorem. But a lot of interesting and important physical
phenomena is described by non-Abelian gauge theories.
Hence it would be very interesting and also fruitful to
have at our disposal a non-Abelian version of the Stokes
theorem. Since non-Abelian differential forms need
different treatment, one is forced to use a more
sophisticated formalism to deal with this new situation.
The aim of this chapter is to present a version of the
non-Abelian Stokes theorem in the framework of the
{\elevenit path-integral} formalism [2].

\vglue 0.6cm
{\elevenbf\noindent 2. From Stokes theorem to Stokes theorem}
\setcounter{currentnumber}{2}
\vglue 0.4cm

The (Abelian) Stokes theorem says that 
we can convert an integral around a closed curve
$\cal C$ bounding some surface $\cal S$ into an integral
defined on this surface. Namely,
$$
\oint_{\cal C} \vec{A} \cdot d\vec{s}
=\int_{\cal S} \mbox{curl}\vec{A} \cdot
\vec{n} d\sigma , 
\eqno{\eqn{1}}
$$
where the curve $\cal C$ is the boundary 
of the surface $\cal S$, i.e. $\cal C=\partial \cal S$,
$\vec{A}$ is a vector field, e.g.\ the vector 
potential of electromagnetic field,
and $\vec{n}$ is a unit outward normal at 
the area element $d\sigma$.
More generally, in any dimension,
$$
\int_{\partial \cal M} \omega=\int_{\cal M} d\omega ,
\eqno{\eqn{2}}
$$
where $\cal M$ is a $d$-dimensional manifold,
$\partial \cal M$ its $(d-1)$-dimensional 
boundary, $\omega $ is a $(d-1)$-form,
and $d\omega $ is its differential, a $d$-form.
We can also rewrite Eq.~2.1 in the spirit of Eq.~2.2, i.e.
$$
\int_{\partial \cal S=\cal C} A_i dx^i=
\frac {1}{2} \int_{\cal S}
\left( \partial_i A_j-\partial_j A_i \right) dx^i \wedge
dx^j , 
\eqno{\eqn{3}}
$$
where $A_i$ ($i=1,2,3$) are components of $\vec A$, 
and the Einstein summation convention should be applied.

In electrodynamics, we define the stress
tensor of electromagnetic field
$$
F_{ij}=\partial_i A_j -\partial_j A_i , 
\eqno{\eqn{4}}
$$
and the magnetic induction, its dual, as
$$
B_k=\frac {1}{2} \varepsilon_{ijk}F_{ij}, 
$$
where $\varepsilon_{ijk}$ is the totally antisymmetric
(pseudo-)tensor. 
RHS of Eq.~2.3 represents then the magnetic flux.
In turn, in geometry $\vec A$ plays the role of  connection 
(it defines the parallel transport around $\cal C$), and $F$ is
the curvature of this connection.

Unfortunately, it is not possible to mechanically 
generalize the Abelian Stokes theorem (Eq.~2.3) to the
non-Abelian one. In the non-Abelian case one faces a
qualitatively 
different situation because
the integrands assume  values in  a Lie
algebra  $\bf g$ rather 
than in  the field of complex numbers  $\bf  C$.
The picture simplifies significantly 
if one switches from the infinitesimal language to a global
one. 
Therefore let us consider the holonomy around 
a closed curve $\cal C$,
$$
{\rm Hol}_{\cal C}(\vec A)=\exp 
\left( i\oint_{\cal C} A_i dx^i \right) . 
\eqno{\eqn{5}}
$$
The holonomy, Eq.~2.5, represents a parallel-transport
operator 
around $\cal C$ assuming values in the Abelian Lie group
${\rm U}(1)$, the gauge group of electromagnetic
interactions. 
Interestingly, the holonomy has a physical meaning
(it is a gauge-invariant object playing the role of the
phase which can be observed in the Aharonov-Bohm
experiment), whereas $\vec A$ has not.

If $\cal C=\partial \cal S$ in Eq.~2.5 we obtain 
a {\elevenit global} version of the Abelian Stokes theorem
$$
\exp \left( i \int_{\partial \cal S=\cal C}
A_i(x) dx^i \right) =
\exp \left( \frac {i}{2} 
\int_{\cal S} F_{ij}(x) dx^i \wedge dx^j \right),
\eqno{\eqn{6}}
$$
which is rather a trivial generalization of  Eq.~2.3.
But nevertheless Eq.~2.6 is a good starting point 
for our further discussion 
concerning the non-Abelian Stokes theorem.

For our further convenience let us formulate
an auxiliary ``Schr\" odinger problem''
governing LHS of Eq.~2.6, 
$$
i\frac {d\psi}{d\tau}=-\dot{x}^i A_i \psi ,
\eqno{\eqn{7}}
$$
which expresses the fact that  the ``wave function'' $\psi$
should be covariantly constant along~$\cal C$,
$$
D_\tau \psi \equiv \left( \frac {d}{d\tau} -i\dot{x}^i A_i
\right) \psi =0,
$$
where $D_\tau$ is the absolute covariant derivative, 
$\tau$ is a parameter on the curve $\cal C$ which is 
analytically defined
by $x^i(\tau)$, and the dot means differentiation 
with respect to the ``time'' $\tau$.

Now, we would like to remind the reader of the  form
of the (``classical'') {\elevenit operator} version of
the non-Abelian Stokes theorem [3]. 
The assumed conventions are as follows. The non-Abelian
curvature or the field strength is defined by
$$
F_{ij}=\partial_i A_j-\partial_jA_i-i[A_i,A_j],
$$
where the connection or the gauge potential assuming values
in an 
irreducible representation $R$ of the Lie algebra $\bf g$
is of the form
$$
A_i=A_i^a T^a,\qquad T^{a\dag}=T^a, 
$$
where the Hermitian generators $T^a=T^a_{rs}$,
$r,s=1,2,\ldots,\dim R$ fulfil the commutation relation
$$
[T^a,T^b]=if^{abc}T^c.
$$
The
non-Abelian generalization of Eq.~2.6 reads
$$
{\rm P} \exp \left( i\oint_{\partial \cal S=\cal C}
A_i(x) dx^i \right)
= {\cal P} \exp \left( \frac{i}{2}
\int_{\cal S} {\cal F}_{ij} (x) dx^i\wedge dx^j \right),
\eqno{\eqn{8}}
$$
where $\rm P$ denotes path ordering, and $\cal P$
some ``surface ordering'' (see Ref.~3, for details).
Here, ${\cal F}_{ij} (x)$ is a 
``path-dependent curvature'' defined by the formula
$$
{\cal F}_{ij} (x) \equiv U^{-1}(x,O) F_{ij}(x) U(x,O),
$$
where $U(x,O)$ is a parallel-transport operator along
the path $\ell$ in the surface $\cal S$ joining 
the base point $O$ of $\partial \cal S$ with the point
$x$, i.e.
$$
U(x,O)={\rm P} \exp \left( i\int_{\ell}
A_i(y)dy^i \right) .
$$

\indent 

In the same manner as quantum mechanics, initially
formulated in the operator language and next reformulated
into the path-integral one, we can translate the operator
form of the non-Abelian Stokes theorem into the
path-integral one. It appears that the key object of the
approach is a two-dimensional {\elevenit topological
quantum field theory} (with a large ``topological''
symmetry) in an external gauge field.  In order to
formulate the non-Abelian Stokes theorem in the
path-integral language we will make the following three
steps:
\begin{enumerate}

\item We will derive a holomorphic path-integral 
representation for the parallel-trans\-port operator
calculating the transition amplitude between one-particle
states of some auxiliary Hilbert-space problem (a
path-integral counterpart of LHS in Eq.~2.8);

\item We will quantize a two-dimensional topological
quantum field theory in an external gauge field $A$
(a counterpart of  RHS in Eq.~2.8);

\item We will apply the Abelian
Stokes theorem.
\end{enumerate}

\vglue 0.6cm
{\elevenbf\noindent 3. The parallel-transport operator}
\setcounter{currentnumber}{3}
\vglue 0.4cm

First of all, let us derive the path-integral
expression for the parallel-transport operator $U$
along an arc of the curve $\cal C$. To this end, 
we should consider 
the non-Abelian formula (differential equation) 
analogous to the Abelian  Eq.~2.7,
$$
i\frac {d\psi_{rs}}{d\tau}=-\dot{x}^i(\tau ) A_i^a
(x(\tau)) T^a_{rs}\psi_s ,
\eqno{\eqn{1}}
$$
where $\psi$ is a ``wave function'' in the irreducible
representation $R$ of the gauge Lie group $G$, which is
to be parallelly transported along the arc of $\cal C$ 
parametrized by $x^i(\tau )$, 
$\tau '\le \tau \le \tau ''$. Formally, Eq.~3.1 can
be instantaneously integrated out yielding
$$
\psi_r (x'')=U_{rs}(x'',x')\psi_s (x') ,
$$
where $x''=x(\tau '')$, $x'=x(\tau ')$, and
$$
U(x'',x')\equiv U(\tau '', \tau ')=
{\rm P} \exp \left( i\int_{\tau '}^{\tau ''}
\dot{x}^i (\tau ) A_i(x(\tau )) d\tau \right) .
\eqno{\eqn{2}}
$$
\indent Let us  now consider  the following auxiliary
mechanical problem with the classical
Lagrangian
$$
L(\bar{\psi}, \psi )=i \bar{\psi} D_\tau \psi ,
\eqno{\eqn{3}}
$$
where now $D_\tau\equiv\frac{d}{d\tau}-i\dot x^iA^a_iT^a$.
The equation of motion following from Eq.~3.3  reproduces
Eq.~3.1, and yields the classical Hamiltonian
$$
H=i\dot{x}^i(\tau )A_i^a(x(\tau )) 
T_{rs}^a \pi_r \psi_s
=-\dot{x}^i (\tau )A_i^a(x(\tau ))T_{rs}^a
\bar{\psi}_r\psi_s .
\eqno{\eqn{4}}
$$
\indent The corresponding auxiliary quantum-mechanical
problem is given,  
according to Eq.~3.4, by the Schr\" odinger equation
$$
i\frac {d}{d\tau} | \Phi \rangle =\hat{H}(\tau)|\Phi
\rangle , 
$$
$$
\hat{H}(\tau )  =-\dot{x}^i(\tau ) A_i^a(x(\tau ))
T_{rs}^a\hat{a}_r^+\hat{a}_s 
$$
$$
\equiv H_{rs}(\tau )
\hat{a}_r^+ \hat{a}_s 
$$
$$
\equiv -\dot{x}^i (t)A_i^a (x(t))\hat{T}^a ,
\eqno{\eqn{5}}
$$
where the creation and anihilation operators satisfy the
standard commutation ($-$) or anticommutation ($+$)
relations
$$
[\hat{a}_r ,\hat{a}_s^+]_{\mp}= \delta_{rs} ,\quad
[\hat{a}_r^+ ,\hat{a}_s^+ ]_{\mp}
=[\hat{a}_r ,\hat{a}_s ]_{\mp} =0 .
\eqno{\eqn{6}}
$$
\indent It can be easily checked by direct computation 
that we have obtained a realization of the Lie algebra
$\bf g$ in a Hilbert space,
$$
[ \hat{T}^a,\hat{T}^b ]_-
=i f^{abc}\hat{T}^c ,
\eqno{\eqn{7}}
$$
where $\hat{T}^a=T_{rs}^a \hat{a}_r^+\hat{a}_s$. For the 
irreducible representation $R$
the second-order Casimir operator $C_2$ is proportional
to the identity operator $\bf 1$, which in turn, is equal
to the number operator $\hat{N}$ in our Fock representation,
i.e.\ if $T^a\rightarrow \hat{T}^a$, then ${\bf
1}\rightarrow \hat{N} 
=\delta_{rs} \hat{a}_r^+ \hat{a}_s$. Thus, by virtue of Eq.~3.7,
we obtain an important for our further considerations constant
of motion $\hat{N}$,
$$
[\hat{N},\hat{H} ]_- =0 .
\eqno{\eqn{8}}
$$
\indent Let us now derive the holomorphic path-integral
representation 
for the kernel,
$$
U(\bar{a},a;\tau '', \tau ')
=\left< \bar{a} \left| {\rm P} \exp \left\{ -i\int_{\tau
'}^{\tau ''} 
\hat{H}(\tau ) d\tau \right\} \right| a \right> ,
$$
of the evolution operator $U$. Literally repeating the
standard textbook procedure [4] (it should be noted that our
approach is in the spirit of the ``physical approach''
to the index theorem, see e.g.\ [5]), we obtain
$$
U(\bar{a},a;\tau '', \tau ')
=\int \exp \left\{ \bar{a}(\tau '')a(\tau '')
+i \int_{\tau '}^{\tau ''} L(\bar{a}(\tau), a(\tau))d\tau
\right\} 
D\bar{a} Da ,
\eqno{\eqn{9}}
$$
where $L(\bar{a}(\tau),a(\tau))$ is the classical 
Lagrangian of the form given in Eq.~3.3,
the imposed boundary conditions are:
$\bar{a} (\tau '')=\bar{a}$, $a(\tau ')=a$, and 
$D\bar aDa$ is a functional ``measure''.
Depending 
on the  statistics  (Eq.~3.6) there are the two ($\mp $)
possibilities
$$
a_r\bar{a}_s \mp \bar{a}_s a_r
=\bar{a}_r \bar{a}_s \mp \bar{a}_s \bar{a}_r
=a_r a_s\mp a_s a_r =0 ,
$$
equivalent as far as one-particle subspace of the Fock
space is concerned, which takes place 
in our further considerations.

Let us confine our attention to the one-particle
subspace of the Fock space. As the number operator $\hat{N}$
is conserved by  virtue of Eq.~3.8, if we start from the
one-particle subspace of the Fock space, we shall remain in
this subspace during all the evolution. The transition
amplitude $U_{rs}(\tau '',\tau ')$ between the one-particle
states $| 1_r\rangle =\hat{a}_r^+ | 0\rangle$ and
$| 1_s\rangle = \hat{a}_s^+| 0\rangle$ is given by the
following scalar product in the holomorphic representation 
$$
U_{rs}(\tau '',\tau ')=\int 
U(\bar{a},\alpha ;\tau '',\tau ')e^{-\bar{a} a}
e^{-\bar{\alpha} \alpha}a_r \bar{\alpha}_s
d\bar{a} da d\bar{\alpha} d\alpha .
\eqno{\eqn{10}}
$$
One can easily check that Eq.~3.10 represents the object
we are looking for. Namely, from the Schr\"odinger
equation (Eq.~3.5) it follows that for the general 
one-particle state $\alpha_r\hat{a}_r^+ | 0\rangle$
(summation after repeating indices) we have
$$
i\frac{d}{d\tau} (\alpha_p \hat{a}_p^+| 0\rangle )
=H_{rs}(\tau )\hat{a}_r^+ \hat{a}_s \alpha_p \hat{a}_p^+
| 0\rangle
=H_{rs}(\tau )\alpha_s (\hat{a}_r^+ | 0\rangle ). 
\eqno{\eqn{11}}
$$
Using the property of linear independence of 
Fock-space vectors in Eq.~3.11, and comparing Eq.~3.11
to Eq.~3.1, we can see that Eq.~3.10 really
represents the matrix elements of the
parallel-transport operator.
According to Eq.~3.9, we can finally put Eq.~3.10 in the
following path-integral form:
$$
U_{rs}(\tau '',\tau ')
=\int \exp \left\{ -\bar{a}(\tau ')a(\tau ')
+i\int_{\tau '}^{\tau ''} L(\bar{a}(\tau),a(\tau)) d\tau
\right\} 
a_r(\tau '') \bar{a}_s(\tau ') D\bar{a} Da ,
\eqno{\eqn{12}}
$$ where the free boundary conditions are imposed.  (One
can also rewrite Eq.~3.12 in the symmetrized form [4].) For
closed paths, $x(t')=x(t'')=x$, Eq.~3.12 gives the holonomy
operator $U_{rs}(x)$. Since $U_{rr}$ is the famous Wilson
loop, it seems that this formula could have some
independent applications.  Interestingly enough, the Wilson
loop, which is supposed to describe a quark-antiquark
interaction, is represented by a ``true'' quark and
antiquark field, $a$ and $\bar{a}$, respectively.  So, the
mathematical trick can be interpreted physically.
Obviously, the ``full'' trace of the kernel in Eq.~3.9 is
obtained by imposing (anti-)periodic boundary conditions in
the case of (anti-)commuting fields, and integrating with
respect to all the variables without the boundary term.
Analogously, one can also derive the parallel-transport
operator (a generalization of the one just considered) for
symmetric $n$-tensors (bosonic $n$-particle states) and for
$n$-forms (fermionic $n$-particle states).

\vglue 0.6cm
{\elevenbf\noindent 4. The non-Abelian Stokes theorem}
\setcounter{currentnumber}{4}
\vglue 0.4cm

Let us now define a (boson or fermion) Euclidean
two-dimensional {\elevenit topological} field theory 
of the fields $\bar \psi,\psi$ in the irreducible
representation $R$ of the Lie algebra $\bf g$
on the compact
surface $\cal S$, $\mbox{dim} {\cal S}=2$, 
$\partial {\cal S}\neq \emptyset$, $\cal S\subset \cal M$,
$\mbox{dim}{\cal M}=d$, in the external gauge field $A$ by
the classical action
$$
S_{\rm{cl}}=\int \left( iD_i \bar{\psi}D_j\psi +\frac{1}{2}
\bar{\psi} F_{ij}\psi \right) dx^i\wedge dx^j,
\qquad i,j=1,\ldots ,d,
\eqno{\eqn{1a}}
$$
or in the parametrization $x^i(\sigma^1,\sigma^2)$,
$\tau '\le \sigma^1 \equiv \tau \le \tau ''$,
$0\le \sigma^2\le 1$, by the action
$$
S_{\rm{cl}}  =
\int_{\cal S} {\cal L}_{\rm{cl}} (\bar{\psi}, \psi )d^2
\sigma 
$$
$$
=\int_{\cal S} \varepsilon^{AB}
\left( i D_A\bar{\psi} D_B\psi +\frac{1}{2} \bar{\psi}
F_{AB} 
\psi \right) d^2\sigma ,\quad A,B=1,2,
\eqno{\eqn{1b}}
$$
where
$$
D_A=\partial_A x^iD_i, \qquad F_{AB}=\partial_A x^i
\partial_B x^j F_{ij}.
$$
The described theory possesses the following
``topological'' gauge symmetry:
$$
\delta \psi (x)=\theta (x), \qquad
\delta \bar{\psi} (x)=\bar{\theta} (x),
\eqno{\eqn{2}}
$$
where $\theta (x)$ and $\bar{\theta }(x)$ are
arbitrary except at the boundary $\partial \cal S$
where they vanish. The origin of the symmetry (Eq.~4.2)
will become clear when we convert the action (Eq.~4.1)
into a line integral. Integrating by parts in Eq.~4.1
and using the Abelian Stokes theorem we obtain
$$
S_{\rm{cl}}=i\oint_{\partial \cal S}
\bar{\psi}D_i\psi dx^i,
\eqno{\eqn{3a}}
$$
or in the parametrization
$$
S_{\rm{cl}}=i\oint_{\partial \cal S} \bar{\psi}
D_{\tau} \psi d\tau .
\eqno{\eqn{3b}}
$$
To covariantly quantize the theory we shall introduce
the BRS operator $s$. According to the form of the gauge
symmetry 
(Eq.~4.2) the operator $s$ is easily defined by
$$
s\psi =\phi,\qquad s\bar{\psi}=\bar{\chi}, \qquad
s\phi =0, \qquad s\bar{\chi}=0,
$$
$$
s\bar{\phi}=\bar{\beta}, \qquad s\chi =\beta , \qquad
s\bar{\beta}=0,\qquad s\beta =0,
$$
where $\phi$ and $\bar{\chi}$ are ghost fields in the 
representation $R$, associated to $\theta$ and $\bar{\theta
}$, 
respectively, $\bar{\phi}$ and $\chi$ are the corresponding
antighosts, and $\bar{\beta}$, $\beta$ are Lagrange
multipliers. 
All the fields possess a suitable Grassmann parity
correlated with the parity of $\bar{\psi}$ and $\psi$.
Obviously $s^2=0$, and we can gauge fix the action in
Eq.~4.1 
in a BRS-invariant manner by simply adding the following
$s$-exact term:
$$
S'  = s\left( \int_{\cal S} \left( \bar{\phi}\triangle \psi
\pm \bar{\psi}\triangle \chi\right) d^2 \sigma \right) 
\eqno{\eqn{4}}
$$
$$
=\int_{\cal S} \left( \bar{\beta} \triangle \psi
\pm \bar{\phi} \triangle \phi \pm \bar{\chi} \triangle \chi
+\bar{\psi} \triangle \beta \right) d^2 \sigma .
$$
The upper (lower) sign stands for the fields $\bar{\psi}$,
$\psi$ of boson (fermion) statistics. Integration after
the ghost fields yields only some numerical factor and
the quantum action
$$
S=S_{\rm{cl}}+\int_{\cal S} \left( \bar{\beta}
\triangle \psi +\bar{\psi} \triangle \beta \right)
d^2\sigma.
\eqno{\eqn{5}}
$$
If necessary, one can insert $\sqrt{g}$ into the second
term, which is equivalent to  change of variables.
Thus the partition function is given by
$$
Z=\int e^{iS}D\bar{\psi} D\psi D\bar{\beta} D\beta ,
\eqno{\eqn{6a}}
$$
with the boundary conditions: $\bar{\beta}|_{\partial \cal S}=
\beta |_{\partial \cal S}=0$. 

One can observe that the job
the fields $\bar{\beta}$ and $\beta$ are supposed to do
consists in eliminating a redundant integration inside
$\cal S$.
The gauge-fixing condition following from Eq.~4.5 imposes
the following constraints 
$$
\triangle \psi =0, \qquad \triangle \bar{\psi}=0. 
$$
Since  values of the fields $\psi $ and $\bar{\psi}$ are
fixed on the boundary $\partial \cal S=\cal C$, we deal
with  two
well-defined $2$-dimensional Dirichlet problems. 
The solutions of the Dirichlet problems fix values of
$\psi$ and $\bar\psi$ inside $\cal S$.
Another, more singular gauge-fixing, is proposed in [2].

Accordingly,
we can rewrite Eq.~4.6a in the form
$$
Z=\int e^{iS_{\rm{cl}}} \left( D\bar{\psi} D\psi \right)
\left|_{\partial \cal S}\right. ,
\eqno{\eqn{6b}}
$$
where the integration is confined to the boundary
$\partial \cal S$.  One can say that, in a sense, we have
{\elevenit BRS-quantized} the Abelian Stokes theorem
passing from the theorem formulated for the classical
action to the theorem formulated for the partition function
(path integral). One can observe that by virtue of the
Abelian Stokes theorem for a closed curve $\cal C$, $\cal
C=\partial \cal S$, Eq.~4.6b is essentially equivalent to
Eq.~3.12, modulo some boundary terms.  It appears that this
``quantized'' Abelian Stokes theorem is a prototype of our
main theorem.

At present, we are prepared to formulate a holomorphic
path-integral version of the non-Abelian Stokes theorem.
Strictly speaking, a particular version of this theorem
(actually, the best-known one) that is applicable to the
case of the Lie algebra valued $1$$(2)$-form, i.e.
connection (curvature) on $1$$(2)$-dimensional space.

In the parametrized form the theorem reads
$$
\int \exp \left[ -\bar{a}(\tau ')a(\tau ')
+i\int_{\tau '}^{\tau ''} L(\bar{a}(\tau), a(\tau)) d\tau
\right] 
a_r(\tau '')\bar{a}_s(\tau ')D\bar{a}Da 
$$
$$
=\int \exp \left[ -\bar{a}(\tau ',0) a(\tau ',0)
+i\int_0^1 \int_{\tau '}^{\tau ''}
{\cal L}_{\rm{cl}}(\bar{a},a) d\tau d\sigma^2 \right]
a_r(\tau '', 0)\bar{a}_s(\tau ',0) D\bar{a} Da,
\eqno{\eqn{7}}
$$ where $L(a,\bar{a})$ and ${\cal L}_{\rm{cl}}(\bar{a},a)$
are defined by Eq.~3.3 and Eq.~4.1b respectively. The
measure on both sides of Eq.~4.7 is the same, i.e. it is
concentrated on the boundary $\partial\cal S$ as in
Eq.~4.6b, and the imposed boundary conditions are free.
Eq.~4.7 can also be put in a parametrization-free form
using Eq.~4.1a and Eq.~4.3a.  By virtue of our earlier
analysis, the proof of the Eq.~4.7 is an immediate
consequence of the Abelian Stokes theorem applied to
$S_{\rm{cl}}$, whereas it follows from the holomorphic
path-integral representation of the parallel-transport
operator that the LHS of Eq.~4.7 really represents the LHS
of the ``operator'' version of the non-Abelian Stokes
theorem (Eq.~2.8). It should be noted that the surface
integral on the RHS of Eq.~4.7 depends on the curvature $F$
as well as on the connection $A$ entering the covariant
derivatives, which is reminiscent of the path dependence of
the curvature $\cal F$.

\vglue 0.6cm
{\elevenbf\noindent 5. Final remarks}
\setcounter{currentnumber}{5}
\vglue 0.4cm

In this final section we shall discuss the two issues:
further generalizations of the presented version of the 
non-Abelian Stokes theorem and its possible physical
applications. 

The proposed theorem is a very particular, though seemingly
the most important, non-Abelian version of the Stokes
theorem. It connects a differential $1$-form in dimension
$1$ and $2$-form in dimension $2$.  The forms are of a very
particular shape, namely, the connection $1$-form and the
curvature $2$-form. Thus, possible generalizations should
concern arbitrary differential forms in arbitrary
dimensions. Since there might be a lot of variants of such
a theorem depending on a particular mathematical and/or
physical application, we will only confine ourselves to
presenting a general recipe.

The general idea is very simple. First of all, we should
construct a topological field theory of auxiliary
topological fields on $\partial \cal M$, the boundary of
the manifold $\cal M$, in the external gauge field we are
interested in. Next, we should quantize the theory, i.e.\
build the partition function in the form of a
path-integral, where auxiliary topological fields are
properly integrated out. Applying the Abelian Stokes
theorem to the (effective) action (in the exponent of the
integrand) we obtain the ``RHS'' of the non-Abelian Stokes
theorem. If we also wish to extend the functional measure
to the whole $\cal M$ we should additionaly quantize the
theory to eliminate the redundant functional integration
inside $\cal M$.

The main source of applications of the non-Abelian Stokes
theorem is coming from topological field theory of
Chern-Simons type.  Path-integral procedure gives the
possibility of obtaining skein relations for knot and link
invariants. In particular, it appears that only the
path-integral version of the non-Abelian Stokes theorem
permits us to nonperturbatively and covariantly generalize
the method of obtaining topological invariants [6].

As a by-product of our approach we have computed the
parallel-transport operator $U$ in the holomorphic
path-integral representation (see Eq.~3.12). In this way,
we have solved the problem of saturation of Lie-algebra
indices in the generators $T^a$. This issue appears, for
example, in the context of equation of motion for
Chern-Simons theory in the presence of Wilson lines (see
Ref.~7, where an interesting connection with the
Borel-Weil-Bott theorem and quantum groups has also been
suggested). Our approach enables us to write those
equations in terms of $\bar{a}$ and $a$ purely classically.
Incidentally, in the presence of Chern-Simons interactions
the auxiliary fields $\bar{a}$ and $a$ acquire fractional
statistics, which could be detected by braiding.  To
determine the braiding matrix one should, in turn, find the
so-called monodromy matrix, e.g. making use of non-Abelian
Stokes theorem.

\vglue 0.6cm
{\elevenbf\noindent 6. Acknowledgments}
\setcounter{currentnumber}{6}
\vglue 0.4cm
The author is grateful to
Professors P. Kosi\'nski, J. Rembieli\'nski, A. Ushveri\-dze
and to Mr.\ H. D\pla browski for interesting discussions.
The work has been partially supported by the KBN grants
2P30213906 and 2P30221706p01.

\vglue 0.6cm
{\elevenbf\noindent References}
\vglue 0.4cm

\bigskip\bigskip\bigskip
%Addendum

{\tt
ADDENDUM

The non-Abelian Stokes theorem was first worked out in
M.~B.~Halpern, Phys.\ Rev.\ D19 (1979) 517, where it was
also used to develop field strength and dual variable
formulations of gauge theory. As an extension to this
development, the theorem was also used on the lattice
by Batrouni a few years later.}

\end{document}